\documentstyle[12pt,aps]{revtex}

\def\centereps#1#2#3{\vskip#2\relax\centerline{\hbox to#1{\special
  {eps:#3 x=#1, y=#2}\hfil}}}

\def\centerbmp#1#2#3{\vskip#2\relax\centerline{\hbox to#1{\special
  {bmp:#3 x=#1, y=#2}\hfil}}}

\def\be{\begin{equation}}
\def\ee{\end{equation}}
\def\bea{\begin{eqnarray}}
\def\eea{\end{eqnarray}}

\begin{document}

\title{\baselineskip14pt SPINORS, RELATIVITY AND
NONLOCALITY\thanks{Some parts of this work
have also been presented at the seminar
at the Brigham Young University, Provo, USA.
August 29, 2000.}}

\author{{\bf VALERI V. DVOEGLAZOV}}

\address{\baselineskip14pt Universidad de Zacatecas,
Apartado Postal 636, Suc. UAZ\\ Zacatecas 98062 Zac., M\'exico\\
E-mail: valeri@ahobon.reduaz.mx\\
URL: http://ahobon.reduaz.mx/\~~valeri/valeri.htm
}

\date{Received December 31, 2000}

\maketitle

\begin{abstract}
\baselineskip14pt
{\bf Abstract.} The Ryder relation between left- and right-
spinors has been generalized in my previous works.
On this basis Ahluwalia presented
a physical content following from this generalization.
It is related to {\it non-locality}. A similar conclusion
can be drawn on the basis of a generalization of Sakurai-Gersten
consideration.  I correct several calculating and conceptual
misunderstandings of the previous works.
\end{abstract}

\newpage

\baselineskip15pt

On the basis of the consideration~\cite{Novozh,Faust,Ryder}
of the Lorentz transformations of the form:
\begin{mathletters}
\begin{eqnarray}
&& x^\prime = \frac{x+vt}{(1-v^2/c^2)^{{1\over 2}}}, \,
y^\prime = y,\, z^\prime =z\,,\\ && t^\prime =
\frac{t+vx/c^2}{(1-v^2/c^2)^{{1\over 2}}} \end{eqnarray}
\end{mathletters}
and rotations in three-dimensional space
\begin{equation}
r^\prime = R r,\quad R^T R = I
\end{equation}
one can obtain the $4\times 4$ matrix representation of the boost and
rotation generators. They form the Lorentz algebra:
\begin{mathletters}
\begin{eqnarray}
&&\left [ K_x , K_y \right ] =
-iJ_z
\quad\text{and\,\,cyclic\,\,permutations}\,,\label{a1}\\
&&\left [ J_x , K_x \right ]
= 0 \quad\text{etc.}\,,\label{a2}\\
&&\left [ J_x , K_y \right ] =
iK_z \quad\text{and\,\,cyclic\,\,permutations}\,,\label{a3}\\
&& \left [ J_x, J_y \right ]
= iJ_z \quad\text{and\,\,cyclic\,\,permutations}\,.\label{a4}
\end{eqnarray}
\end{mathletters}
It was shown that the Lorentz transformations are connected with
the squeeze transformations~\cite{K1}. In the $(1/2,0)\oplus (0,1/2)$
representation it was explicitly shown that the Lorentz group
is essentially $SU(2)\otimes SU(2)$, ref~\cite[p.40]{Ryder}.

The Relativity Theory conserves the {\it interval}, $ds^2 = dx^\mu
dx_\mu$\,, $\mu = 0, 1, 2, 3$. As a consequence, the second-order
differential equation follows:
\begin{equation}
({1\over c^2} {\partial^2 \over \partial t^2} - \vec \nabla^2 )
\phi + {m^2 c^2 \over \hbar^2} \phi = 0\label{kg}
\end{equation}
for a field without spin splitting. The Dirac equation hence appears as a
relation between 2-spinors of the spinor representations $({1\over 2},0)$
and $(0,{1\over 2})$ of the algebra
(\ref{a1}-\ref{a4}).\footnote{ The helicity eigenspinors can be
parametrized as follows~\cite[p.180]{Var},\cite{DV1,Itzyk}:
\begin{mathletters}
\begin{eqnarray} &&\xi_\uparrow =
Ne^{i\vartheta_1} \pmatrix{\cos(\theta/2)\cr \sin (\theta/2)
e^{i\phi}\cr}\,,\\ &&\nonumber\\
&&\xi_\downarrow = Ne^{i\vartheta_2}\pmatrix{\sin
(\theta/2)\cr -\cos (\theta/2) e^{i\phi}\cr}\,,
\end{eqnarray}
\end{mathletters}
with  $\theta$ and $\phi$, the angles of $\vec p$ in the spherical
coordinate system; $N$ and $\vartheta_{1,2}$ are arbitrary parameters.}
Of course, its solutions satisfy
the momentum-space realization of (\ref{kg}).
Under parity the representations interchange
$(j,0) \leftrightarrow (0,j)$, because matrices of the Lorentz
transformations (rotations and boosts) for dotted and undotted spinors are
related by the Wigner operator (see the equation (2.75) in~\cite{Ryder})
\begin{equation} \Lambda_{_L}= \zeta \Lambda_{_R}^\ast
\zeta^{-1},\quad\mbox{with}\,\, \zeta = -i\sigma_2\,.
\end{equation}
Ryder writes:``Now when
a particle is at rest, one cannot define its spin
as either left- or right-handed, so $\phi_{_R} (\vec 0)=\phi_{_L}
(\vec 0)$.
It then follows ... that"
\begin{mathletters}
\begin{eqnarray}
\phi_{_R} (\vec p) \,&=&\, \Lambda_{_R} (\vec p \leftarrow
\vec 0)\, \phi_{_R} (\vec 0) =
\Lambda_{_R}  (\vec p \leftarrow \vec 0)
\,\phi_{_L} (\vec 0) =\nonumber            \\
&=&\Lambda_{_R} (\vec p
\leftarrow \vec 0) \,\Lambda_{_L}^{-1} (\vec p\leftarrow
\vec 0)\,\phi_{_L} (\vec p)\,, \label{eq1} \\
&&\nonumber\\
\phi_{_L} (\vec p) \,&=&\, \Lambda_{_L} (\vec p \leftarrow
\vec 0) \,\phi_{_L} (\vec 0) \, =
\Lambda_{_L} (\vec p \leftarrow \vec 0) \,
\phi_{_R} (\vec 0 ) =\nonumber\\ &=&
\Lambda_{_L} (\vec p \leftarrow \vec 0) \,\Lambda_{_R}^{-1}
(\vec p \leftarrow \vec 0) \,\phi_{_R}
(\vec p)\,,\label{eq2}
\end{eqnarray}
\end{mathletters}
where boosts were only used.
In the $4\times 4$ matrix form (after the corresponding substitutions
of quantum-mechanical operators $E\rightarrow i\hbar {\partial \over
\partial t}$, $\vec p \rightarrow -i\hbar \vec\nabla$ and $c=\hbar =1$)
the equations (\ref{eq1},\ref{eq2}) become to be written
\begin{equation} [i\gamma^\mu
\partial_\mu - m ] \psi (x) = 0\,.  \end{equation}
This derivation of the Dirac equation has been analyzed in~\cite{ajp}.
Different ways of derivations of  the Dirac equation (
its generalizations and  higher spin equations) have been presented
in~\cite{DVN} with corresponding citations.

However, the declaration of impossibility to distinguish spin of a
particle as
either left- or right-handed assumes that it is also possible to set
\begin{equation}
\phi_{_R} (\vec 0) = e^{i\alpha} \phi_{_L} (\vec
0)\,,
\end{equation}
with arbitrary parameter $\alpha$. This has been
studied in~\cite{DV1,DV2,DV3,DV4,DV5}, see also~\cite{Pashkov,DVA-R} and
the commented  paper~\cite{DVA}. Furthermore,
the relation of the Ryder book can
be generalized in different ways, see the generalized
formulas (8) and (10), e.~g., in ref.~[11a] and the formulae (5)
in~\cite{DV5}:
\begin{equation} \phi_{_L}^h
(\overcirc{p}^\mu) = a (-1)^{{1\over 2} - h} e^{i(\vartheta_1
+\vartheta_2)} \Theta_{[1/2]} [\phi_{_L}^{-h} (\overcirc{p}^\mu)]^\ast + b
e^{2i\vartheta_h} \Xi^{-1}_{[1/2]} [\phi_{_L}^h (\overcirc{p}^\mu)]^\ast
\,
\end{equation}
(the notation is explained in the cited papers).
Our intention of
modification of the Dirac formalism originates from the classical works of
Markov, Gelfand and Tsetlin, and Sokolik~\cite{Mar-Sok}.  Recently, by
using the coordinate-dependent phase Ahluwalia derived the `CP-violating
Dirac equation', ref.~\cite{Jap}, and suggested to interpret  the resulting
non-locality as that which ``manifests in the spinorial space (i.e., the
spinorial indices) and not in the configurational space (i.e., the $\vec
x$ space)"~\cite{DVA}.   Unfortunately, I have to correct several
Ahluwalia's claims.  I show that this interpretation is incorrect: the
resulting non-locality manifests itself in the coordinate space in the
sense that the fermionic anticommutator does {\it not} vanish outside the
light cone $(x-x^\prime)^2 <0$ (cf. the discussion in
ref.~\cite[p.150]{Itzyk}).\footnote{I am grateful to an anonymous referee
of {\it Physical Review D} on the paper~[17d]
``Additional Equations Derived from
the Ryder Postulates in the $(1/2,0)\oplus (0,1/2)$ Representation of
the Lorentz Group.' for independent
confirmation of this obvious fact. I acknowledge discussions of the first
version of the AFDB paper with Dr.  Ahluwalia in the beginning of 1998
during his second visit in Zacatecas, M\'exico.} This Ahluwalia's
misconception originates from a calculating error in the derivation of the
formulas (23,24) of ref.~\cite{DVA}.\footnote{It is not
very convenient for a reader to have different
physical quantities denoted by the same symbol (as in~\cite[Eq. (4)]{DVA}
for 2-spinors and phases). But, in order not to mislead those who have
read the work~\cite{DVA} and is reading this comment, we shall follow the
notation of the commented paper.}
I also give several remarks on the Ahluwalia
misunderstandings, provide several insights into the problem of the
``kinematical" CP violation and present relations with
research of other authors.

First of all, let me check the correctness of the derivation of the
formula (23).
It is very strange from the first sight that in
the formula (23) the left-hand side depends only on $x^\mu$ and
$x^{\mu^\prime}$, but the right-hand side depends on the
$k^0=\sqrt{{\vec k}^{\,2} +m^2}$,
see the  formula
(24) for ${\cal O}_{ij}$.  Even if one assumes that one can neglect the
difference of coordinate-dependent phases in $u$ (and $v$) 4-spinors in
points $\vec x$ and $\vec x^{\,\prime}$  for which the anticommutator
(23) has been calculated (see the footnote $d$ in~\cite{DVA}), we note
that Ahluwalia put the term ${1\over k_0}$ outside the integration on $d^3
{\vec k}$.\footnote{One should remember
that the transformation $d^4 k \,\delta (k^2 -m^2) \sim {d^3 \vec k \over
2k_0}$ is used when transferred to the 3-dimensional momentum-space
integrals in the field operators. Furthermore, one can also show that
the connection between $k_0$ and $\vec k$ is used when deriving the
Dirac-(like) equations by the Ryder method, see~\cite[Eq.(10)]{DVA}.} We
give here corrected calculations in detail.  The equal-time anticommutator
is:  \begin{eqnarray} && \{\Psi_i (\vec x, t), \Psi_j^\dagger (\vec
x^\prime,t)\}_+ = \sum_{\sigma\sigma^\prime} \int \int {d^3 \vec k \, d^3
\vec k^\prime \over (2\pi)^6} {m^2 \over k_0 k_0^\prime} \left [u^\sigma_i
(\vec k) \overline u^{\sigma^\prime}_k (\vec k^\prime) \gamma^0_{kj}
\{b_\sigma , b_{\sigma^\prime}^\dagger\} e^{-ikx + ik^\prime
x^\prime}+\right .\nonumber\\
&&\left . + v^\sigma_i (\vec k) \overline
v^{\sigma^\prime}_k (\vec k^\prime) \gamma^0_{kj} \{d_\sigma^\dagger ,
d_{\sigma^\prime}\} e^{ikx - ik^\prime x^\prime}\right ] = \sum_{\sigma}
\int {d^3 \vec k \over (2\pi)^3} {m \over k_0} \left [u^\sigma_i (k)
\overline u^{\sigma}_k (k) \gamma^0_{kj} e^{+i\vec k (\vec x - \vec
x^\prime)}+ \right.\nonumber\\
&&\left.
v^\sigma_i (k) \overline v^{\sigma}_k (k) \gamma^0_{kj}
e^{-i\vec k (\vec x - \vec x^\prime)}\right ] =\nonumber\\ && =  {1\over
\cos (\phi)} \int {d^3 \vec k \over (2\pi)^3} {m \over k_0} \left
[(\widehat k +\zeta_u^{-1} m)_{ik} \gamma^0_{kj} e^{i\vec k (\vec x - \vec
x^\prime)}+ (\widehat k +\zeta_v^{-1} m )_{ik} \gamma^0_{kj} e^{-i\vec k
(\vec x - \vec x^\prime)}\right ] =\nonumber\\ && =  {1\over \cos (\phi)}
\int {d^3 \vec k \over (2\pi)^3} {m \over k_0} \left [2k_0 \delta_{ij} +m
(\zeta_u^{-1} +\zeta_v^{-1})_{ik} \gamma^0_{kj} \right ] e^{i\vec k (\vec
x - \vec x^\prime)} = \nonumber\\ && = {2m \over \cos (\phi)} \left [
\delta_{ij} \delta (\vec x -\vec x^\prime )  +im \sin (\phi)
(\gamma^5 \gamma^0)_{ij} D^1 (0, \vec x -\vec x^\prime )
\right ]\, ,\label{ac}
\end{eqnarray}
where $D^1 (x^\mu -x^{\mu^\prime})$ is the even solution of
the homogeneous Klein-Gordon equation.
Its explicit form was given in ref.~\cite[formula (A2B.6)]{Bogoliubov}
(see also p. 150 of~\cite{Itzyk}):
\begin{eqnarray}
D^1 (x) &=& i (D^+ (x) - D^- (x)) = \\
&=& {m\over 4\pi \sqrt{\lambda}}
\theta (\lambda) N_1 (m\sqrt{\lambda}) + {m \over 2\pi^2 \sqrt{-\lambda}}
\theta (-\lambda) K_1 (m\sqrt{\lambda}) \approx\nonumber\\
&-&{1\over 2\pi^2 \lambda} +{m^2 \over 4\pi^2}
ln {m\sqrt{\vert\lambda\vert} \over 2}\, ,\nonumber
\end{eqnarray}
with $\lambda =(x^0)^2 -\vec  x^{\,2}$, and $N_1$, $K_1$ are
the first-order cylinder functions (the Neumann function and
the MacDonald-Hankel function, respectively).  As it
is readily seen, the
formula (\ref{ac}) does {\bf not} coincide with the formula (23)
of~\cite{DVA}.

Let me also present the result of calculation of the
anticommutator of two free fields at arbitrary separations,
the analogue of the Pauli-Jordan function for this kind of $(1/2,0)\oplus
(0,1/2)$ fields:
\begin{eqnarray} \{ \Psi (t, \vec x), \overline{\Psi} (t^\prime
,\vec x^\prime ) \}_+ &=& {2m \over i\cos (\phi)} [ i\widehat\partial_x
+m\cos (\phi) ] D (x -x^\prime) + \nonumber\\
&+& 2im^2 \gamma^5 \tan
(\phi) D^1 (x-x^\prime)\, .\label{pp} \end{eqnarray}
Since the function
$D^1 (x-x^\prime) \neq 0$ for $(x-x^\prime)^2 <0$, the `local'
observables would not commute at equal times (cf. with the condition
(1.5) of~\cite{Wein}).\footnote{One can still write the equation
(\ref{pp}) in a symbolic form with the operator $\hat\epsilon=
i\partial_t/\vert i\partial_t\vert$ (introduced by Weaver, Hammer and
Good, Jr.~\cite{Weav})):  \begin{equation} \{ \Psi (t, \vec x),
\overline{\Psi} (t^\prime ,\vec x^\prime ) \}_+ = {2m \over i\cos (\phi)}
[ i\widehat\partial_x +m\cos (\phi) + im \gamma^5 \sin (\phi)\hat \epsilon
] D  (x-x^\prime) \, .\label{pp1} \end{equation}
It is relevant to the
dynamical equation obtained~\cite{DV1,DV2,Jap,DVA} in the approximation
$\phi (x) \approx const$:
\begin{equation} [ i\gamma^\mu \partial_\mu -
m\cos (\phi) \pm im\gamma^5 \sin (\phi) ] \Psi_\pm (x^\mu) =0\,
.\label{en} \end{equation}
But, in my opinion, such a formulation only put
cover on the intrinsic non-locality of the theory in the $\vec x$-space.
Of course, the last term in (\ref{en}) can be considered as
a Higgs-like `interaction'.  We noted that equations
with $\gamma^5$ `interaction' term have also been introduced (apart of
our previous works) in the context of the Dirac oscillator~\cite{Dixit}
and of the Dirac supersymmetry and pseudoscalar-Higgs mass
generation~\cite{Moreno}.} In conclusion, the derived non-locality is {\it
not} the non-locality, which ``manifests in the spinorial space"; the
anticommutators explicitly contains the even solution $D^1 (x-x^\prime)$
of the Klein-Gordon equation in the formulas (\ref{ac},\ref{pp}), which
does not vanish for space-like intervals $(x-x^\prime)^2 < 0$.
The non-locality is obviously in the $\vec x$-space.  The contribution of
the even $D^1 (x-x^\prime)$ function is larger for small intervals, and if
$t-t^\prime =0$ we have the inverse proportionality to $\vert\Delta \vec
x\vert^2$.  In a more detailed version of this paper (submitted to {\it
Mod. Phys. Lett. A}) we took into account the dependence of
phase factors on the space points and instead of Eqs. (21) and (22) of the
commented paper, the analogues of the projection operators, we obtained
generalized expressions, as well as those instead of Eqs. (23) and (24)
of~\cite{DVA}.

I want to indicate that  the concept of a {\it variable}
coordinate-dependent mass is {\it not} a new one, e. g.~\cite{vari}
(see also old related papers~\cite{nonl}).  Moreover, these results can
be obtained {\it without} the use of the Ryder procedure, {\it but},
instead, one can use the Sakurai-Gersten method~\cite{SG}. I start from
\begin{equation}
(E^2 -c^2 \vec{p}^{\,2}) I^{(2)}\Psi =
\left [E I^{(2)} + c \vec{p}\cdot \vec{\sigma} \right ]
\left [E I^{(2)} - c \vec{p}\cdot \vec{\sigma} \right ]
\Psi = M^2 c^4 \Psi \label{G1}
\end{equation}
(cf. Eq. (4) of~[26b])\,.
Then, its solutions can be found from
\begin{mathletters} \begin{eqnarray} (i\hbar {\partial \over \partial
x^0} +i\hbar \vec{\sigma}\cdot \vec{\nabla}) \Psi (x) &=&m(x) \,c\Phi
(x)\,,\\
(i\hbar {\partial \over \partial x^0}  - i\hbar
\vec{\sigma}\cdot \vec{\nabla} ) m(x) c\Phi (x) &=& M^2 c^2 \Psi (x)\,.
\end{eqnarray} \end{mathletters}
or
\begin{mathletters} \begin{eqnarray}
(i\hbar {\partial \over \partial
x^0} +i\hbar \vec{\sigma}\cdot \vec{\nabla}) \Psi (x)&=&Mc \phi(x) \Phi
(x) \,,\label{ff1}\\
Mc \{ \phi (x) (i\hbar {\partial \over \partial x^0}
- i\hbar \vec{\sigma}\cdot \vec{\nabla} ) \Phi (x) &+& (i\hbar {\partial
\phi(x)\over \partial x^0} -i\hbar (\vec{\sigma}\cdot \vec{\nabla}
\phi(x) ) \Phi (x)\}= \nonumber\\
&&\qquad\qquad\qquad = M^2 c^2 \Psi \,.\label{ff2} \end{eqnarray}
\end{mathletters}
In the 4-component form after algebraic transformations
one has
\begin{eqnarray} \lefteqn{\pmatrix{-{Mc\over \phi(x)}& i\hbar {\partial
\over \partial x^0} - i\hbar \vec{\sigma}\cdot \vec{\nabla}\cr i\hbar
{\partial \over \partial x^0} + i\hbar \vec{\sigma}\cdot \vec{\nabla}&
-Mc \phi(x))\cr} \pmatrix{\Psi\cr \Phi\cr} +}\nonumber\\
&&+ \pmatrix{0&{1\over \phi(x)}
(i\hbar {\partial \phi(x)\over \partial x^0} -i\hbar \vec{\sigma}\cdot
\vec{\nabla}\phi(x))\cr 0&0\cr}\pmatrix{\Psi\cr \Phi\cr} = 0\,.
\end{eqnarray}
So, we have an equation \begin{equation} [ i\hbar
\gamma^\mu \partial_\mu - {Mc\over \phi(x)} {1+\gamma^5 \over 2}-
Mc\phi(x) {1-\gamma^5 \over 2} + i\hbar \gamma^\mu_- {1\over \phi(x)}
\partial_\mu \phi(x) ] \psi^D (x) = 0\,.  \end{equation}
When presenting
$\phi(x) = \exp (\eta + i\chi (x))$ we obtain:
\begin{equation} [i\hbar
\gamma^\mu \partial_\mu - Mce^{-\eta-i\chi(x)} {1+\gamma^5 \over 2} -
Mce^{\eta+i\chi(x)} {1-\gamma^5 \over 2} -\hbar \gamma^\mu_-
\partial_\mu \chi (x)] \psi^D (x)= 0\,.  \end{equation}
The
charge-conjugate Dirac field function satisfies the equation with the
opposite sign before some terms (according to the  formulas
in~\cite[footnote 3]{acta}).  You may see that these
equations lead to a theory, which is similar to that based on Eq.
(\ref{en}). As opposed to that, the formulation based on the
Sakurai-Gersten procedure is a manifestly relativistic covariant one.
In general, one can consider the right-hand side of
(\ref{ff1}) to be $Mc (\sigma^\mu c_\mu) \Phi$ or $Mc \sigma^2 \sigma^\mu
\tilde c_\mu \Phi^\ast$, or even in more general form (like
in~\cite{DV1}).  Additional terms in the Dirac-like equation will answer
to some specific forms of interaction.

Finally, I present several remarks.

{\bf Remark 1} to the Section 2.1 of the paper~\cite{DVA}
follows.\footnote{Below the formulas numeration refers to the Ahluwalia
paper.} That  author introduces the $\lambda$ and $\rho$ spinors
\begin{equation}
\lambda(p^\mu)\,\equiv \left( \begin{array}{c}
\left(\zeta_\lambda\,\Theta_{[j]}\right)\,\phi^\ast_{_L}(p^\mu)\\
\phi_{_L}(p^\mu)\\
\end{array}
\right)\,\,,\quad
\rho(p^\mu)\,\equiv
\left(
\begin{array}{c}
\phi_{_R}(p^\mu)\\
\left(\zeta_\rho\,\Theta_{[j]}\right)^\ast\,\phi^\ast_{_R}(p^\mu)
\end{array}
\right)
\,\,.\label{os}
\end{equation}
with $\Theta_{[j]}$ being the Wigner time-reversal operator and he
writes: ``... for fermion fields these phases must take on the
values $\pm i$ to ensure that the spinors of the $(j,0)\oplus (0,j)$
representation are self/anti-self charge conjugate, i.e. they
are of the extended Majorana type."

The spin-1/2 charge-conjugate operator, which was defined in old papers,
is:
\begin{equation}
S^c_{[1/2]} =e^{i\vartheta^c_{[1/2]}} \pmatrix{0 & i\Theta_{[1/2]}\cr
-i\Theta_{[1/2]} & 0\cr}{\cal K}\, .\label{def}
\end{equation}
As readily seen from the condition of self/anti-self conjugacy,
the phases $\zeta_{\lambda,\rho}$
(which Ahluwalia refers to) depend on the phase factor in the definition
(\ref{def}):
\begin{equation}
\zeta_\lambda = \pm ie^{+i\vartheta^c_{[1/2]}},\,\,\,
\zeta_\rho = \pm ie^{-i\vartheta^c_{[1/2]}}\, .
\end{equation}
For instance, if $\vartheta^c_{[1/2]} = \pi/2$ then one has
$\zeta_\lambda = \mp 1$ and $\zeta_\rho = \pm 1$.

Presumably, the same result will hold for higher fermion spins.

{\bf Remark 2.} The 8-component Dirac-like equations (i.~e.
obtained on the basis of the different
choice of phase factors between left- and right- momentum-space 2-spinors)
have been given (apart from~\cite{DV3} and references therein)
in~[16a] and references therein. They naturally lead to the idea of
opposite gravitational masses of particle and its antiparticle (cf.  the
Introduction in ref.~\cite{DVA} and the Santilli results~\cite{Sant}
obtained in different frameworks).

{\bf Remark 3.} It was recently shown~\cite{DVO-EV} that additional
phase factors in the definition of parts of the field operator may lead
(apart of difficulties in constructing a scalar Lagrangian in the case
under consideration\footnote{For instance, I am not aware of any proof if the term
${\cal L}\sim \overline \Psi \hat\epsilon \Psi$ ($\hat \epsilon$
is the Weaver-Hammer-Good sign operator~\cite{Weav}) would be
a scalar in any case of definition of the field operator.}) to unusual
relativistic transformation laws of the corresponding Noether  currents.

{\bf Remark 4.}\footnote{I am very grateful to an anonymous
referee of {\it Foundation of Physics} for the discussion on
how is the dimension of field operators to be fixed.
If we would not be careful in this question some paradoxes
related to the Noether currents eigenvalues may arise.}
The dimension of spin-1/2 fermion field operator is usually
chosen to be equal to $[energy]^{3/2}$ in the system of units $c=\hbar =1$.
This is the consequence of the convention that
the action used in the variation procedure must be {\it dimensionless}
and, hence, the Lagrangian density must have the dimension $[energy]^4$.
Thus, taking into account the definitions of classical 4-spinors (16,17)
we conclude that the creation/annihilation operators used in Eq. (19)
of the cited paper
should have the dimension $[energy]^{-2}$.
In the mean time the Ahluwalia equation (20), the anticommutation
relations, manifests that the creation/annihilation operators have the
dimension $[energy]^{-3/2}$; let us not forget that the dimension of the
delta-function is inverse to its argument.  Therefore,  we either have to
add some mass factor to the denominator of the right-hand side of (20) or
to substitute the mass factor $m$ by $\sqrt{m}$ in the definition of the
field operator Eq. (19) of~\cite{DVA}. Finally, one can also present the
mass term in the Lagrangian for spin-1/2 field in a rather unusual form,
${\cal L}\sim \overline\Psi \Psi$.
It is the latter case which was implicitly implied in Ahluwalia's paper
(we learnt this looking at the dimension of the right-hand side of his
equation (23)).

{\bf Remark 5.} The basis of (16,17) is not much convenient
because in the case of $\phi (x)={\pi\over 2} (2n+1)$, $n=0,1,2\ldots$ we
would have divergent behavior of the corresponding 4-spinors. Reasons of
the choice of this basis have not been given therein.

{\bf Remark 6.} Another possibility for construction of the corresponding
4-spinors exists:  to use the first $\pm$ sign in~\cite[Eq.(4)]{DVA} for
spin-up and spin-down spinors (and not for $\phi_{_{R,L}}$
spinors), respectively.

{\bf Remark 7.} With respect to the Ahluwalia's footnote $f$, C.
DeWitt-Morette {\it et al.} have already argued that the sign of metric
may have physical significance~\cite{DeWitt}.

On the basis of the above, our conclusion is the following: though the
Ahluwalia physical idea is interesting, its presentation suffers from
several misunderstandings and it contains many encrypted statements.
Mathematical and physical foundations of this theory have been presented
in several papers before.  In the present article the non-locality in
$\vec x$-space was explicitly shown, which may lead to the violation of
the Causality Principle (or even to the energy-momentum non-conservation).
However, I agree that this formalism deserves further elaboration,
particularly in the way, if the usual momentum-space commutation relations
(used in~\cite[Eq.(20)]{DVA}) may be considered to be valid in these
frameworks.

\acknowledgments
I greatly appreciate discussions with M. Berrondo,
A. E. Chubykalo, J.-F. Van Huele, Y. S. Kim, S. Kruglov, I. Nefedov, A. F. Pashkov, Z.
Silagadze and Yu.~F.~Smirnov. Information from and discussions with D.
Ahluwalia (1993-1998) are acknowledged. This work has been partly
supported by the Mexican Sistema Nacional de Investigadores.

\end{document}